\documentclass[prl,showpacs,twocolumn,amsmath,amssymb]{revtex4-1} 

\usepackage{graphicx}

\usepackage[nooneline,raggedright]{subfigure}

\usepackage{hhline}

\usepackage{bm} 

\usepackage[dvips]{color}

\newcommand{\eref}[1]{Eq.~(\ref{#1})}
\newcommand{\fref}[1]{Fig.~\ref{#1}}

\newcommand{\tref}[1]{Tab.~\ref{#1}}

\newcommand{\im}{%
           \imath}

\newcommand{\up}{%
        \ensuremath{\uparrow}}
\newcommand{\dn}{%
        \ensuremath{\downarrow}}

\newcommand{\com}[1]{}

\renewcommand\Re{\hbox{Re}}
\renewcommand\Im{\hbox{Im}}

\newcommand{\vek}[1]{%
        \hbox{\textbf #1}}
\newcommand{\svek}{%
        \mathbf}

\newcommand{\out}[1]{}

\def\XXint#1#2#3{{\setbox0=\hbox{$#1{#2#3}{\int}$}
\vcenter{\hbox{$#2#3$}}\kern-.5\wd0}}

\begin{document}

\title{On the separability of dynamical and non-local correlations in three dimensions}  

\author{T. Sch\"afer}
\author{A. Toschi}
\author{Jan M. Tomczak}
\email{corresponding author; jan.tomczak@tuwien.ac.at}
\affiliation{Institute of Solid State Physics, Vienna University of Technology, A-1040 Vienna, Austria}

\begin{abstract}
While second-order phase transitions always cause strong non-local fluctuations, their effect on spectral properties
crucially depends on the dimensionality.
For the important case of three dimensions, we show that the electron self-energy is well {\it separable} into a local dynamical
part and static non-local contributions.
In particular, our non-perturbative many-body calculations for the 3D Hubbard model at different fillings demonstrate
that the quasi-particle weight remains essentially momentum-independent, also in the presence of overall large
non-local corrections to the self-energy.
Relying on this insight we propose a ``space-time-separated'' scheme for many-body perturbation theory that is up to ten times more efficient
than current implementations.
Besides these far-reaching implications for state-of-the-art 
electronic structure schemes, our analysis will also provide guidance to the quest of going beyond them.
\end{abstract}




\maketitle

\paragraph{Introduction.}
Several iconic phenomena of the many-body problem, such as the Kondo effect or the Mott metal-insulator transition,
can be described by {\it local} correlation effects. 
This explains the great success of dynamical mean-field theory\cite{bible} (DMFT)
for
our understanding of numerous correlated materials. 
However, 
DMFT 
{\it ad hoc} assumes the electron self-energy to be independent of momentum.
This is known to fail in low dimensions,
e.g.\ for the Luttinger liquid in 1D, or the strong momentum space differentiation 
in (quasi) 2D systems.
However, even in three dimensions --the major realm of practical DMFT applications-- signatures of non-local spatial correlations
are apparent, 
e.g.,
in the presence of second order phase transitions:
In the 3D Hubbard model, nearest-neighbor spin-spin-correlation functions\cite{PhysRevLett.106.030401,PhysRevB.87.205102}, non-mean-field critical exponents\cite{PhysRevLett.107.256402}, 
and deviations from a local correlations' picture of the entropy\cite{PhysRevLett.106.030401,PhysRevB.87.205102} indicate
a paramount effect of non-local antiferromagnetic fluctuations 
in a large region of the phase-diagram.

Similarly, 
for realistic correlated materials,  
thought to be well described by the combination, DFT+DMFT\cite{RevModPhys.78.865}, of density functional theory with DMFT, 
important non-local exchange and correlation effects have recently been established\cite{jmt_pnict,jmt_svo,PhysRevB.87.115110,PhysRevB.89.235119,jmt_svo_extended}
within the so-called {\it GW} approximation\cite{hedin}
 -- a many-body perturbation theory\cite{RevModPhys.74.601}.
Non-local effects have, for example, been held accountable for a proper description of the Fermi surfaces in the iron pnictides BaFe$_2$As$_2$\cite{jmt_pnict,PhysRevLett.110.167002} 
and LiFeAs\cite{jmt_pnict,PhysRevLett.105.067002}, 
and for the non-magnetic nature of BaCo$_2$As$_2$\cite{paris_sex}.

Complementary to 
these manifestations of self-energy effects that are non-local in {\it  space}, one 
might also investigate
their structure 
in the {\it time} domain.
While exchange contributions to the electron self-energy are static by construction, correlation effects are {\it a priori} both momentum- and energy-dependent.
Recently it has been proposed that
the quasi-particle weight $Z_{\svek{k}}=(1-\partial_\omega\Re\Sigma(\vek{k},\omega))^{-1}_{\omega=0}$, accounting for the low-energy dynamics in the (retarded) self-energy $\Sigma$ of metals, is essentially momentum-independent
in the iron pnictides 
\cite{jmt_pnict}, as well as metallic transition metal oxides\cite{jmt_svo_extended}.

Yet, the basis for the mentioned empirical finding of the locality of $Z_{\svek{k}}$ was the weak-coupling {\it GW} approach, where spin fluctuations are completely neglected.
However, large dynamical spin-fluctuations have been found in the iron pnictides both theoretically\cite{PhysRevB.86.064411,Liu_natphys_pnict} and experimentally\cite{Liu_natphys_pnict}.
Moreover, these fluctuations were shown to constitute the leading contribution to non-local self-energies in the (extended) Hubbard model\cite{ayral_gwdmft,PhysRevLett.107.256402,PhysRevB.87.125149}.

\begin{figure}[!t]
  \begin{center}
{\scalebox{1.}{\includegraphics[angle=-90,width=.45\textwidth]{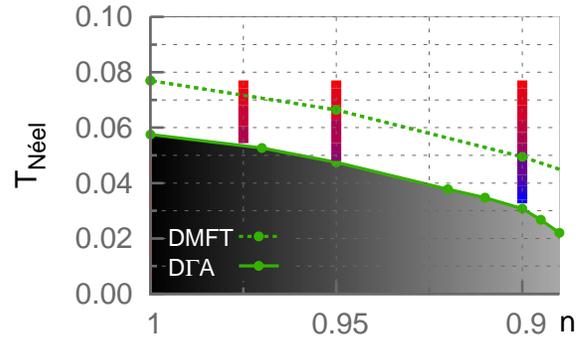}}} 
      \caption{{\bf Phase diagram of the 3D Hubbard model within DMFT and D$\Gamma$A for $U$$=$$1.6$.} 
			The solid (dashed) line indicates the D$\Gamma$A (DMFT) N\'eel temperature, determined from the divergence of the spin susceptibility.	
			The vertical bars at fixed filling $n$ indicate the temperature paths followed in \fref{variance}. The system is Mott-insulating at half-filling ($n=1$). All energies measured in units of the half-bandwidth.}
      \label{phase_diagram}
      \end{center}
\end{figure}

\begin{figure*}[!th]%
\begin{tabular}{ccc}
\multicolumn{3}{l}{$\Sigma(k,\omega)=\quad\quad\pmb{\Re\Sigma(k,\omega=0)}\quad\qquad\qquad+\qquad\qquad\quad\left(1-1/\pmb{Z(k)}\right)\omega\qquad\qquad-\qquad\qquad\quad\im\pmb{\Gamma(k)}\left(\omega^2+\pi^2T^2\right)\quad+\,\cdots$}\\[0.2cm]
  {\scalebox{1.}{\includegraphics[clip=true,trim=10 0 0 10,angle=-90,width=.3\textwidth]{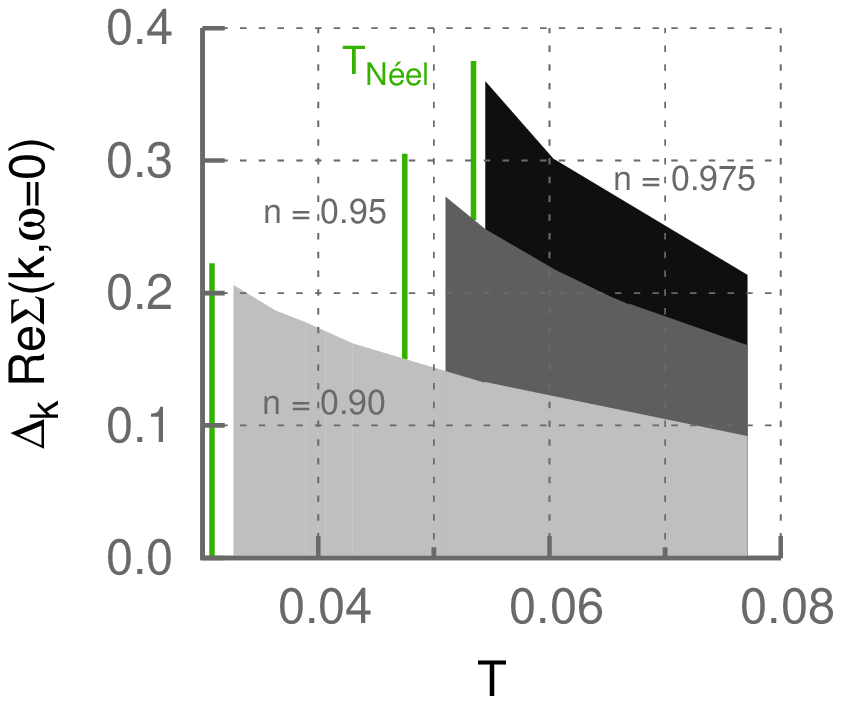}}}   
& {\scalebox{1.}{\includegraphics[clip=true,trim=10 0 0 10,angle=-90,width=.3\textwidth]{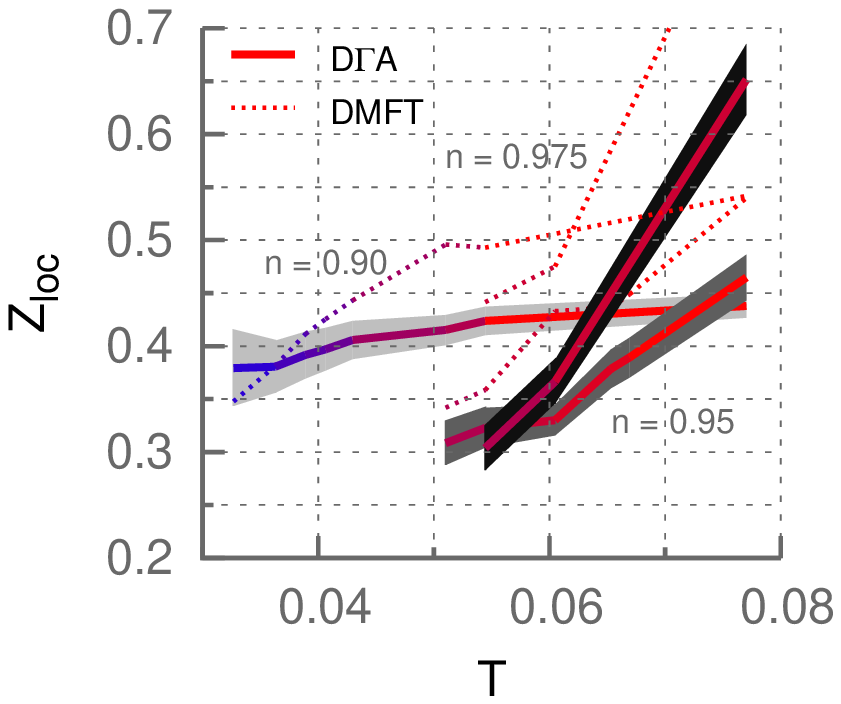}}} 
& {\scalebox{1.}{\includegraphics[clip=true,trim=10 0 0 10,angle=-90,width=.3\textwidth]{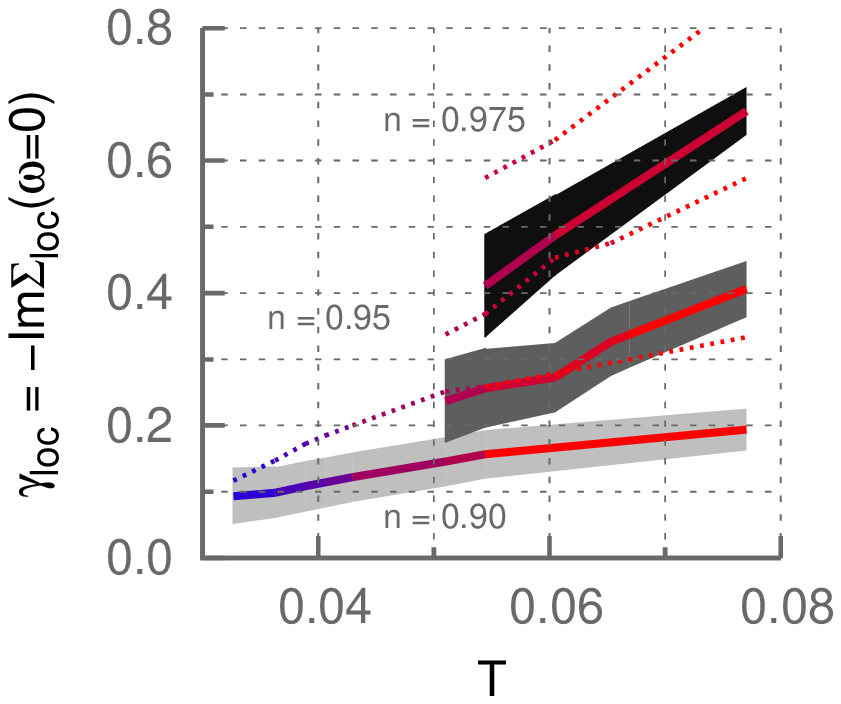}}}
\end{tabular}
\caption{{\bf Low-energy expansion of the D$\Gamma$A self-energy and momentum dependence of the expansion coefficients.} 
The shaded areas (light gray, gray, black) indicate the standard deviation, 
$\Delta_{\svek{k}}a(\vek{k})=\sqrt{1/N_{\svek{k}}{\sum_{\svek{k}}\left| a(\vek{k})-a_{loc} \right|^2}}$, 
of the expansion coefficients $a(\vek{k})=\Re\Sigma(\vek{k},\omega$$=$$0)$, $Z(\vek{k})$, $\gamma(\vek{k})=-\Im\Sigma(\vek{k},\omega$$=$$0)=\Gamma(\vek{k})\pi^2 T^2+\mathcal{O}(T^4)$ in the Brillouin zone with respect to their local values $a_{loc}$, as a function of temperature for different fillings ($n=0.9$, $n=0.95$, $n=0.975$). 
From left to right:
(i) the static real part of the self-energy at the Fermi level, $\Re\Sigma(\vek{k},\omega$$=$$0)$: Its standard deviation increases notably on approaching the N{\'e}el transition (marked by green lines).
The local value of $\Re\Sigma(\vek{k},\omega$$=$$0)$, including the Hartree term, was absorbed into the chemical potential.
(ii) the quasi-particle weight $Z(\vek{k})$, and 
(iii) the scattering rate $\gamma(\vek{k})$. 
For (ii) and (iii) the standard deviation with respect to momentum is shown as stripes (shades of gray) around the respective local value, $Z_{loc}$ and $\gamma_{loc}$,
within D$\Gamma$A (solid lines). As comparison the (by construction) local values of $Z$ and $\gamma$ within DMFT are shown (dotted lines).
$U$$=$$1.6$ and the temperatures and fillings correspond to the vertical cuts shown in \fref{phase_diagram}.
}
\label{variance}
\end{figure*}

Here, we put the analysis of {\it non-local correlations} in spectral properties of metals on solid grounds.
To this aim we apply a diagrammatic extension of DMFT, the dynamical vertex approximation\cite{toschi:045118} (D$\Gamma$A), to the 3D Hubbard model away from half-filling.
This allows for a precise study 
of the electron self-energy {\it beyond} the weak-coupling regime.
We find that while non-local correlation effects increase substantially when approaching the Mott insulating or the
magnetically ordered state, the associated fluctuations 
do not manifest themselves as a sizeable momentum differentiation in the low-energy dynamics of the self-energy.
In particular, the quasi-particle weight is indeed found to be essentially local.
On the other hand, the momentum variation of the {\it static} part of the non-local self-energy reaches a magnitude of 20\% of the half-bandwidth, or more.
Dynamical renormalizations acquire an appreciable momentum dependence only at energies in the outer half of the quasi-particle bandwidth or higher. 
We will discuss the implications of our findings
for electronic structure schemes and make an explicit suggestion that speeds up self-energy calculations for metals within {\it GW}
by a factor of 10.

\paragraph{Model and method.}
Our starting point is the 3D Hubbard model on the cubic lattice,
%
$H=-t\sum_{\left<i,j\right>\sigma}c_{i\sigma}^\dag c_{j\sigma}^{\phantom{\dag}}+U\sum_i n_{i\up}n_{i\dn}$
%
where $c_{i\sigma}^\dag$ ($c_{i\sigma}^{\phantom{\dag}}$) creates (destroys) an electron of spin $\sigma$ at site $i$, $n_{i\sigma}=c_{i\sigma}^\dag c_{i\sigma}^{\phantom{\dag}}$,
$t$ is the hopping amplitude between nearest neighbour sites $\left<i,j\right>$, and $U$ the on-site Hubbard interaction.
Our D$\Gamma$A calculations 
exploit a DMFT input for the local self-energy and vertex functions\cite{PhysRevB.86.125114,PhysRevLett.110.246405}
computed with an exact diagonalization solver  and employ 
the (particle-hole) ladder approximation with Moriya correction in the spin channel\cite{Held01062008,PhysRevB.80.075104}.
Energies will be measured in units of the half-bandwidth $W/2=6t\equiv 1$. 
We choose $U=1.6$, which, at half filling, $n=1$, yields a Mott insulator with maximal N{\'e}el temperature\cite{PhysRevLett.107.256402}.
We thus consider the crossover regime between
weak-coupling (where the perturbative {\it GW} approximation is most justified and magnetism controlled by Fermi surface instabilities) 
and the Mott-Heisenberg physics at large interaction strengths. 
It was shown (for half-filling) that the effect of non-local fluctuations is strongest in this intermediate 
regime\cite{Held01062008,PhysRevB.83.075122,PhysRevB.86.125114}.

\begin{figure*}[!t]
{\includegraphics[clip=true,trim= 20 35 35 40, angle=-90,width=.33\textwidth]{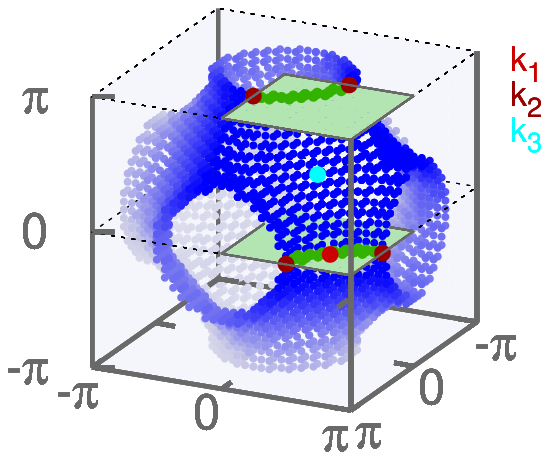}} 
\hspace{-0.25cm} 
{\includegraphics[clip=true,trim= 0 0 20 120, angle=-90,width=.6\textwidth]{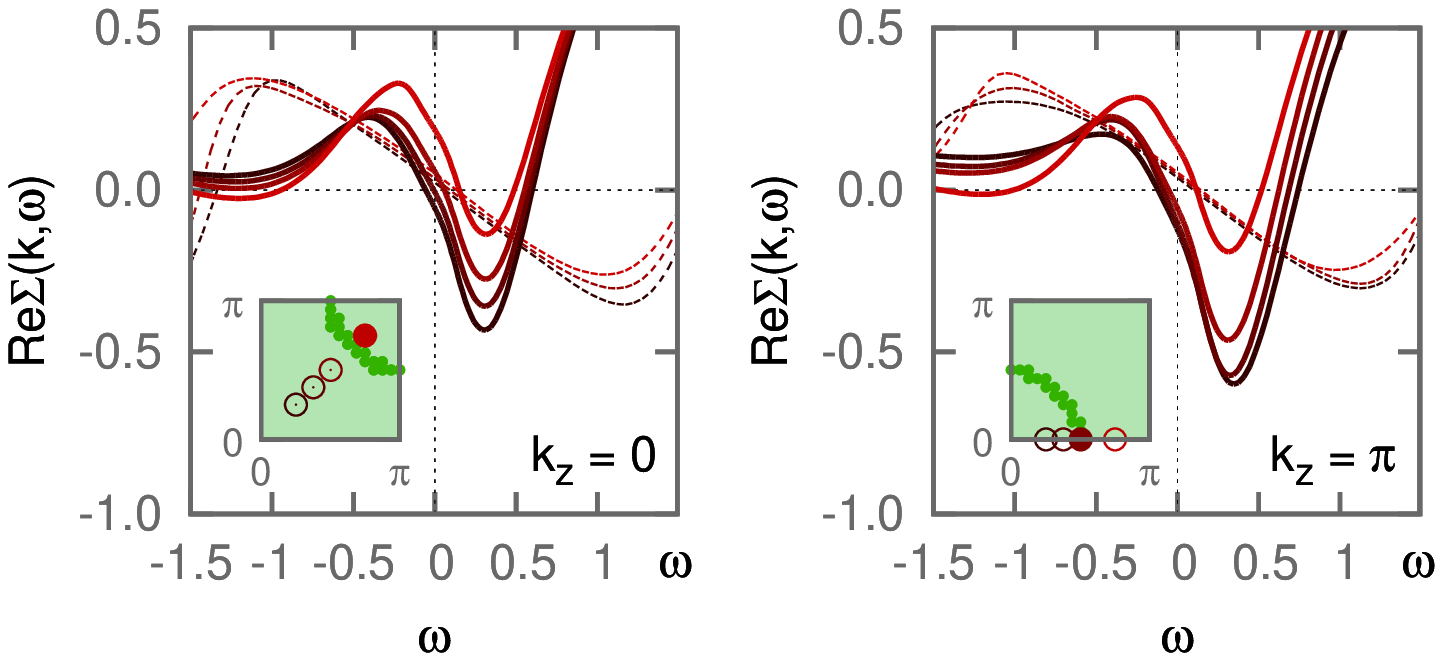}}  
\caption{{\bf D$\Gamma$A Fermi surface and momentum dependence of the self-energy.}  
Shown is the Fermi surface computed in D$\Gamma$A for $n=0.9$, $U=1.6$, $T=0.043$ (left). 
There, the green squares indicate cuts of the Brillouin zone that contain the paths $(k,k,0)$ and $(k,0,\pi)$ for which the real parts of the D$\Gamma$A self-energies (middle \& right panel) are shown. 
Indicated are also the $\vek{k}$-points of \tref{mstar}. The maximal (minimal) $Z$ and $\gamma$ on the Fermi surface occur at $\vek{k}_2$ ($\vek{k}_3$).
We also show self-energy results from the {\it GW} approach (middle \& right, thin dashed lines): {\it GW} overestimates $Z$ and the energetic extent
of the (linear) Fermi liquid regime. In all cases, the constant Hartree contribution was omitted, and the chemical potential is set to zero.
}
\label{FS}
\end{figure*}

\paragraph{Results.}
\fref{phase_diagram} shows the phase diagram of the 3D Hubbard model 
as a function of filling $n$ and temperature $T$.
As a clear signature 
of non-local fluctuations, the N{\'e}el temperature 
is reduced by at least 30\% in D$\Gamma$A with respect to the DMFT result.
To elucidate the influence of these manifestly non-local effects in the {\it two-particle} AF susceptibility
onto the {\it one-particle} electronic structure, we analyse the D$\Gamma$A self-energy when approaching the 
spin density wave (SDW) instability at constant filling.

First, we focus on effects near the Fermi level and perform a low-energy expansion of the self-energy: 
%
$\Sigma(\vek{k},\omega)=\Re\Sigma(\vek{k},\omega$$=$$0)+(1-1/Z(\vek{k}))\omega+\im\Gamma(\vek{k})(\omega^2+\pi^2T^2)+\cdots$,
where $\gamma(\vek{k})=-\Im\Sigma(\vek{k},\omega=0)=\Gamma(\vek{k})\pi^2T^2+\mathcal{O}(T^4)$ is the scattering rate, and $Z(\vek{k})$
can be identified as the quasi-particle weight in the Fermi liquid regime.
We recall that in the limit of infinite dimensions, non-local self-energy diagrams vanish, and $Z$ and $\gamma$ are momentum independent\cite{PhysRevLett.62.324}. The DMFT self-consistency condition then yields via Dyson's equation the exact non-interacting
propagator of an effective Anderson impurity problem\cite{bible}. In finite dimensions this is no longer true. Therefore, besides the approximation of assuming the self-energy to be local,
this local self-energy does not need to coincide with the local projection of the exact lattice self-energy.
In fact, the momentum average $a_{loc}=1/N_\svek{k}\sum_{\svek{k}}a(\vek{k})$ (with $N_\svek{k}$ the number of $\vek{k}$-points) of the D$\Gamma$A quasi-particle weight and scattering rate, $Z_{loc}$ and $\gamma_{loc}$,  deviate notably from the DMFT prediction (\fref{variance}, middle and right panel).
As expected\cite{toschi:045118,PhysRevB.80.075104}, the inclusion of antiferromagnetic fluctuations reduces the quasi-particle weight $Z$.
We note that the temperature evolution of $Z_{loc}$, and its change in hierarchy ($Z$ smallest at low doping, $1-n$, for small $T$; 
while for high $T$, $Z$ is largest for small doping)
follows the same trends as the inverse of the effective mass of the 3D electron gas\cite{PhysRevB.70.035104}.
For the chosen interaction strength, the scattering rate $\gamma$ in DMFT is large enough to induce a large violation of
the pinning condition 
$\Im G_{loc}(\omega=0)=\Im G_{loc}^{U=0}(\omega=0)$, 
valid for local self-energies with vanishing 
imaginary part at the Fermi level\cite{pinning}. 
Moreover, the temperature dependence of $\gamma$ evidently involves corrections\cite{PhysRevB.68.155113} to Fermi liquid theory as 
neither DMFT nor D$\Gamma$A yield a $T^2$ behaviour.
Therewith the interpretation of the expansion coefficient $Z$ as quasi-particle weight breaks down. 
Nevertheless, we shall in the following formally refer to the solutions $\vek{k}_F$ of the quasi-particle equation, $\det (\mu-\epsilon_{\svek{k}}-\Re\Sigma(\vek{k},0))=0$, 
with the chemical potential $\mu$ and the one-particle dispersion $\epsilon_{\svek{k}}$ as ``Fermi surface''\cite{supps}.
%

In D$\Gamma$A, spectral weight at the Fermi level is further depleted compared to DMFT\cite{supps}\cite{toschi:045118,PhysRevLett.107.256402} and effective masses are reduced (see below). 
Hence, the {\it local} electron-electron scattering contributions to quasi-particle lifetimes decrease, explaining why $\gamma^{DMFT}>\gamma_{loc}^{D\Gamma A}$.
When approaching the SDW state, however, non-local spin-fluctuations provide an additional scattering mechanism:
While $\gamma\rightarrow 0$ for $T\rightarrow 0$ in DMFT, the inverse life-time in D$\Gamma$A levels off.
%
Concomitantly, analogous to the incoherence crossover at high-$T$ via local electron-electron scattering, the emerging low-$T$ scattering in D$\Gamma$A,
causes $Z^{D\Gamma A}$ to saturate towards the SDW state. 

We now analyse the momentum dependence of the self-energy by calculating the standard deviation of the 
expansion coefficients $a(\vek{k})=\Re\Sigma(\vek{k},\omega$$=$$0)$, $Z(\vek{k})$, $\gamma(\vek{k})$
with respect to their local values: 
$\Delta_{\svek{k}}a(\vek{k})=\sqrt{1/N_{\svek{k}}{\sum_{\svek{k}}\left| a(\vek{k})-a_{loc} \right|^2}}$
\cite{jmt_pnict,supps}.
We find that non-local fluctuations manifest themselves very differently in the individual coefficients (from left to right in \fref{variance}):

(i) the momentum dependence of the {\it static} part of the self-energy $\Re\Sigma(\vek{k},\omega$$=$$0)$, as measured by the above standard deviation,
increases substantially towards the 
spin ordered phase, and grows sharply when approaching the Mott insulator at half-filling; 
$\Delta_{\svek{k}}\Re\Sigma(\vek{k},\omega=0)$ reaches values as large as 20-40\% of the half-bandwidth $W/2$ --a large effect that is fully neglected in DMFT.

(ii) the standard deviation in momentum space of the quasi-particle weight, $\Delta_{\svek{k}}Z(\vek{k})$ (depicted as shaded areas around the local values in \fref{variance}(b)), is small in
all considered cases. Indeed the largest absolute deviation amounts to only 0.07.
In particular, $\Delta_{\svek{k}}Z(\vek{k})$ does not dramatically increase upon approaching the N{\'e}el temperature, in stark contrast to the discussed static part of the self-energy.

(iii) the momentum dependence of the scattering rate $\gamma$, \fref{variance}(c), remains always moderate.
Specifically, the momentum variation increases on absolute values when approaching the Mott insulator at half-filling,
although the relative importance $\Delta_{\svek{k}}\gamma(\vek{k})/\gamma$ actually decreases.

In all, we thus find that while spin and charge fluctuations, that develop upon approaching the spin-ordered or Mott-insulating state\cite{supps},
can renormalize significantly the value of the quasi-particle weight $Z$, they do not introduce
any sizable momentum differentiation in it.
This is in strong opposition to the pronounced non-local effects in
the static part of the self-energy.
The latter will, however, strongly modify the mass $m^*$ of the quasi-particles, as e.g.\ 
extracted from Shubnikov-de Haas or photoemission experiments.
Indeed the effective mass enhancement $m^*/m$ is defined by the ratio of group velocities of the non-interacting and interacting system, respectively:
\begin{equation}
\left.\left(\frac{m^*}{m^{\phantom{*}}}\right)^{-1}\right|_{\svek{k}_F} = Z(\vek{k}_F) \left[ 1 + \frac{\vek{e}_{{\svek{k}}_F}\cdot\nabla_k \Re\Sigma(\vek{k},\omega=0)}
{\vek{e}_{{\svek{k}}_F}\cdot\nabla_k \epsilon_{\svek{k}}} \right]_{\svek{k}=\svek{k}_F}
\label{mass}
\end{equation}
where $\epsilon_{\svek{k}}$ is the non-interacting dispersion and $\vek{e}_{\svek{k}_F}$ 
is the unit-vector perpendicular to the Fermi surface for a given $\vek{k}_F$.
Thus, besides the enhancement of $m^*$ via $Z$ (which we showed to be quasi local, see also \tref{mstar}), there is a contribution to $m^*$ from the momentum dependence of the static self-energy.
The sign of the derivative $\nabla_k\Re\Sigma$ is always positive, thus the effect of non-local correlations is to {\it reduce} the effective mass.
In \tref{mstar} we give the individual components to $m^*/m$ for three Fermi vectors $\vek{k}_F$ (on the Fermi surface, $Z$ and $\gamma$ are maximal (minimal) for $\vek{k}_2$ ($\vek{k}_3$)).
We find $m^*/m$ to be notably momentum-dependent: $m^*/m=1.4$ for $\vek{k}_1$, while for $\vek{k}_3$ $m^*/m=1.8$ --a value larger by 30\%.
However, 
it is dominantly the {\it spatial} variation of the self-energy ($\nabla_k\Re\Sigma$), 
not a non-local dependence in its dynamics ($Z$), that causes this momentum differentiation.
Depending on $\vek{k}_F$, 
non-local correlation effects reduce the effective mass down to 55-75\% of its dynamical contribution, $1/Z$.
In realistic {\it GW} calculations even larger reductions were found for iron pnictides\cite{jmt_pnict}. 
Besides the change in the (local) quasi-particle weight, this is a second, significant effect not accounted for in local approaches, such as DMFT.

\begin{table}
\begin{tabular}{l|c||c|c|c|c||c}
$\vek{k}=\vek{k}_F$ & $k_0$ &$\nabla_k \epsilon_{\svek{k}}$ & $\nabla_k\Re\Sigma(\vek{k},0)$ & $Z$ &1/$Z$& $m^*/m$\\
\hhline{=|=||=|=|=|=||=}
$\vek{k}_1$$=$$(k_0,k_0,0)$ & 2.16    & 0.55 & 0.30 & 0.45 & 2.20 & 1.42\\
$\vek{k}_2$$=$$(k_0,0,\pi)$ & $\pi/2$ & 0.67 & 0.20 & 0.45 & 2.23 & 1.72 \\
$\vek{k}_3$$=$$(k_0,k_0,k_0)$ & $\pi/2$ & 0.99 & 0.35 & 0.41 & 2.44 & 1.81
\end{tabular}
\caption{{\bf Effective masses on the Fermi surface.} 
Contributions to $m^*/m$ from dynamical ($Z$) and static ($\nabla_k\Re\Sigma(\vek{k},0)$) renormalizations for three $\vek{k}_F$,
see also \fref{FS}. In DMFT: $m^*/m=1/Z=1/0.44=2.26$.
$U$$=$$1.6$, $T$$=$$0.043$. 
}
\label{mstar}
\end{table}

Having so far concentrated on effects at the Fermi level, a natural question is: Up to which energy scale do
dynamical correlations remain essentially local?
\fref{FS} shows the Fermi surface within D$\Gamma$A for $n=0.9$ and the real parts of the self-energies\cite{bla3}
%
following two k-paths in the Brillouin zone. 
Congruent with the quasi-particle weight being quasi local, the slopes of the self-energies at the Fermi level are the same for all momenta
and the curves differ by a static shift only.
This, for the chosen parameters, is approximately true within the window $[-0.4:0.5]$. 
Given the bandwidth renormalization $W\rightarrow Wm/m^*$, with the above effective mass ratio, 
non-local correlations are effectively static over most of the {\it interacting} quasi-particle dispersion.

In \fref{FS} we also show {\it GW} results: While the slope of the self-energy is constant throughout the Brillouin zone within  the linear Fermi liquid regime (the extension of which {\it GW} overestimates), also the static part shows only a weak momentum-dependence%
\cite{bla4}. 
%
The comparatively large variations of $\Re\Sigma(\vek{k},\omega$$=$$0)$ 
in D$\Gamma$A therefore emphasize the pivotal influence of spin fluctuations (neglected by {\it GW})
onto (static) non-local correlations.

\paragraph{Discussion and outlook.}
The central findings of our analysis for correlated metals in 3D are:
(1) Within most of the quasi-particle bandwidth {\it non-local} correlations are static.
Conversely dynamical correlations are {\it local}. Hence, the self-energy of 3D systems is {\it separable} into
non-local and dynamical contributions
\begin{equation}
\Sigma(\vek{k},\omega)=\Sigma^{non-loc}(\vek{k})+\Sigma^{loc}(\omega)
\label{Sep}
\end{equation}
providing an {\it a posteriori} justification for the application of a DMFT-like method to describe $\Sigma^{loc}(\omega)$ in 3D.
We stress however that since e.g.\ $Z^{DMFT}\ne Z^{D\Gamma A}_{loc}$, ways to improve the DMFT impurity propagator (e.g.\ by incorporating
$\Sigma^{non-loc}(\vek{k})$ in the DMFT self-consistency) need to be pursued.
(2)  {\it Static} correlations have a large momentum-dependence,
calling for a description of $\Sigma^{non-loc}$
beyond, say, DFT. 
This can e.g.\ be achieved with the {\it GW}+DMFT approach\cite{PhysRevLett.90.086402,jmt_svo}, or the recently proposed QS{\it GW}+DMFT\cite{jmt_pnict,jmt_sces14}.
Exploiting \eref{Sep}, these can be 
simplified, as is the strategy in DMFT@(non-local {\it GW})\cite{jmt_svo_extended}.
Yet, already the {\it GW} can profit: 
We here propose to replace Hedin's
$\Sigma_{GW}(\vek{k},\omega)=1/N_{\svek{q}}\sum_{\svek{q},\nu}G(\vek{k}+\vek{q},\omega+\nu)W(\vek{q},\nu)$
with
%
$\widetilde{\Sigma}_{GW}(\vek{k},\omega)=\Sigma_{GW}^{loc}(\omega)+\Sigma_{GW}^{non-loc}(\vek{k})$,
%
where
%
$\Sigma_{GW}^{loc}(\omega)$$=$$\sum_{\nu}G^{loc}(\omega+\nu)W^{loc}(\nu)$,
$\Sigma_{GW}^{non-loc}(\vek{k})$$=$$1/N_{\svek{q}}\sum_{\svek{q},\nu}G(\vek{k}+\vek{q},\nu)W(\vek{q},\nu)-\Sigma_{GW}^{loc}(\omega=0)$
%
for {\it GW} calculations of metals.
We shall refer to this physically motivated scheme as ``space-time-separated {\it GW}''.
Avoiding the $\vek{q}$- and $\omega$-convolution, respectively, reduces the numerical expenditure from $N_{\svek{k}}N_{\svek{q}}  \times N_{\omega}N_{\nu}$ to $N_{\omega}N_{\nu} + N_{\svek{k}}N_{\svek{q}}\times N_\nu$, typically gaining more than an order of magnitude\cite{blaL}.
%
If the dominant non-local self-energy derives from {\it exchange} effects, \eref{Sep} holds 
and SEX+DMFT\cite{paris_sex} can be employed. 
In the (one-band) Hubbard model, however, non-local self-energies are not exchange-driven. 
Still, as we have shown, $\Sigma^{non-loc}$ is significant, and in particular beyond a perturbative description {\`a} la {\it GW}.
Consequently, at least in the vicinity of second order phase transitions, a methodology beyond (QS){\it GW}+DMFT 
is required. 
{\it Ab initio} D$\Gamma$A\cite{ANDP:ANDP201100036} or realistic applications of other diagrammatic extensions\cite{rubtsov:033101,PhysRevLett.102.206401,0953-8984-21-43-435604,PhysRevB.88.115112,PhysRevLett.112.196402} of DMFT might provide a framework for this.
That \eref{Sep} holds {\it beyond} weak coupling, however,
nourishes the hope that a much less sophisticated electronic structure methodology can be devised in 3D.

Non-local renormalizations that are dynamical occur in lower dimensions,
as e.g.\ shown theoretically for 2D\cite{PhysRevLett.95.106402,PhysRevB.79.045133,PhysRevLett.112.196402,RevModPhys.77.1027,2014arXiv1405.7250S}.
However, also in 3D, momentum-dependent quasi-particle weights can be generated.
In fact, this is the typical situation in heavy fermion systems below their (lattice) Kondo temperature.
There, the hybridization amplitude for spin singlets between atomic-like $f$-states and conduction electrons
is modulated on the Fermi surface, as it can be rationalized with mean-field techniques\cite{PhysRevB.77.245108}.
Thus, even a local quasi-particle weight of the $f$-states yields a momentum-space anisotropy of $Z$ via
the change in orbital character. 
This effect has also been held responsible for
anisotropies in some Kondo insulators\cite{JPSJ.65.1769}.
Beyond this scenario,  however,
strong inter-site fluctuations in the periodic Anderson model\cite{PhysRevB.84.115105} suggest actual
non-local correlation effects to be of crucial relevance to heavy fermion quantum criticality\cite{Steglich_HF_review,Paschen_HF_review}.
A further source of non-trivial non-local correlation effects in 3D are multi-polar Kondo liquids\cite{PhysRevB.75.144412,PhysRevB.77.125118,pwave-kondo}. 
To elucidate the latter two phenomena, an application of D$\Gamma$A to e.g.\ the periodic Anderson model is called for.

\paragraph{Acknowledgements.}
We acknowledge financial support from the Austrian
Science Fund (FWF) through project No.\ I-597-N16 and the German Research Foundation (DFG) unit FOR 1346.
TS was supported by the FWF through the Doctoral School
“Building Solids for Function” (project ID W1243).
Calculations have been performed on the Vienna Scientific Cluster (VSC).




\onecolumngrid
\appendix


\bigskip
\begin{center}
{\bf On the separability of dynamical and non-local correlations in three dimensions\\
--Supplemental Material--}
\end{center}

\section{Momentum variation when approaching half-filling at constant temperature}

\begin{figure}[!t!h]
  \begin{center}
{\scalebox{1.}{\includegraphics[angle=-90,width=.475\textwidth]{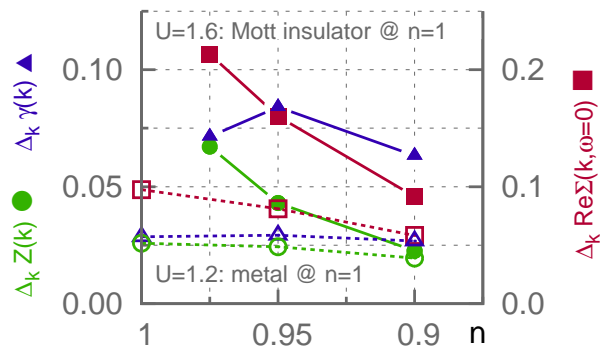}}} 
      \caption{{\bf Standard deviation over the Brillouin zone of the static self-energy $\Re\Sigma(\vek{k},\omega$$=$$0)$, quasi-particle weight $Z$ and the scattering rate $\gamma$ as a function of filling $n$.} 
			The proximity to the Mott insulating state at half-filling for $U$$=$$1.6$ (filled symbols \& solid lines) 
			induces a notably larger momentum differentiation than for $U$$=$$1.2$ (open symbols \& dashed lines) which yields a metal, also at half-filling.
			Contrary to the static self-energy, the overall values for the standard deviation of $Z$ and $\gamma$ remain small (note the different scales).
			$T=0.077$, i.e.\ far above the respective N{\'e}el temperatures. }
      \label{approach_n1}
      \end{center}
\end{figure}

Besides the spin-fluctuations due to the presence of the spin-density wave state, also the proximity to the Mott phase matters
for the existence of non-local self-energies. In \fref{approach_n1} we plot the standard deviations, $\Delta_k$, of the low energy expansion coefficients
comparing two different interaction strengths: $U=1.6$ which is Mott insulating at half-filling (used throughout the main text), and $U=1.2$ which is metallic for all dopings.
The chosen temperature, T=0.077, is reasonably higher than the spin ordering temperatures that are 0.057 and 0.049 at $n=1$ for $U=1.6$ and $U=1.2$, respectively.
The momentum differentiation is significantly more pronounced for the doped Mott insulator than for the system whose parent is a correlated metal.
This is congruent with the finding (at half-filling \cite{PhysRevB.86.125114,Held01062008}) that long-range fluctuations are strongest in the crossover regime ($U\sim 1.6$),
which was our motivation to consider this regime for our main discussion.
In all cases, the dominant momentum dependence resides in the {\it static} part of the self-energy, while $Z$ is essentially local
for all interaction strengths, temperatures and dopings.

\begin{figure}[!t]
  \begin{center}
{\scalebox{1.}{\includegraphics[angle=-90,width=.4\textwidth]{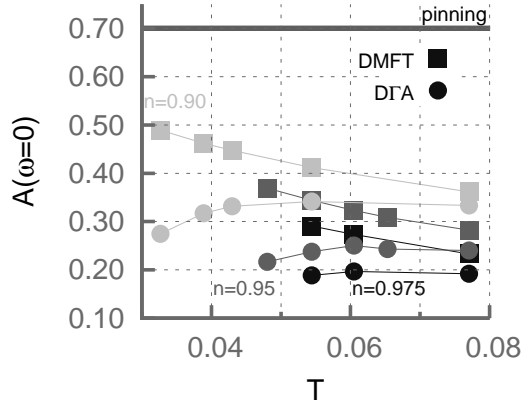}}} 
      \caption{{\bf Spectral weight at the Fermi level.} 
	Shown are the values $A(\omega=0)=-\lim_{i\omega\rightarrow 0}1/\pi \Im G(i\omega)$ within DMFT (squares) and D$\Gamma$A (circles) for different fillings $n$.
  At $n=1$, $U=1.6$ places the system into the crossover regime of the metal-insulator transition. The manifestation of this is the violation of the pinning condition (Luttinger's theorem) for local self-energies
	by the DMFT. In D$\Gamma$A, the spectral weight is further depleted. See text for details.
			} 
      \label{A0}
      \end{center}
\end{figure}

\section{Pinning condition and Luttinger's theorem}

As alluded to in the main text, the crossover regime between the weak-coupling metal and strong coupling Mott insulator at half-filling
is prone to a low quasi-particle coherence temperature.
We recall that for a local self-energy and infinite quasi-particle lifetimes, $\Im\Sigma(\omega=0)=0$, the spectral function at the Fermi level as
obtained e.g.\ from the Matsubara data, $A(\omega=0)=-\lim_{i\omega\rightarrow 0}1/\pi \Im G(i\omega)$, is pinned to its value
in the non-interacting ($U \rightarrow 0$) limit\cite{pinning}.
As is shown in \fref{A0}, this condition is strongly violated within DMFT (where the self-energy is local by construction) for the chosen interaction, temperature and doping conditions, as $\gamma=-\Im\Sigma(i\omega\rightarrow 0)>0.1$ in units of the half-bandwidth. This amounts to a violation of Luttinger's theorem, since a local self-energy cannot change the volume enclosed by the ``Fermi surface''.
Therewith the concepts of a Fermi surface and quasi-particles in general are strictly speaking ill-defined.
As done in Fig.\ 3 of the main text, however, we will define $\vek{k}_F$ as a vector for which $G(\vek{k}_F,\omega=0)^{-1}=\left[\mu-\epsilon_{\svek{k}_F}-\Re\Sigma(\vek{k}_F,\omega=0)\right]=0$, and the respective solutions form the ``Fermi surface''.

\medskip
\noindent
Going beyond a local self-energy has two effects:

(1) A non-local self-energy ($\Re\Sigma(\vek{k},\omega=0)$) deforms the Fermi surface, and pinning needs not to be verified even if quasi-particles have infinite lifetimes.
See e.g.\ Ref.~\onlinecite{jmt_svo_extended} for a discussion for the example of SrVO$_3$.
In our D$\Gamma$A results, the spectral weight at the Fermi level is further reduced with respect to the DMFT results.
Since away from the vicinity of the SDW transition $\gamma^{D\Gamma A}_{loc}<\gamma^{DMFT}$, the increased violation of pinning is owing to the deformation of 
the surface created by the poles of the Greens function 
that would form the Fermi surface if lifetimes were sufficiently long. This is thus one signature of non-local correlations.
At high temperatures the D$\Gamma$A values of $A(\omega=0)$ approach those of the DMFT (limit of ``single ion physics'' in which $\Sigma$ is local).

(2) At very low temperatures, $\gamma^{D\Gamma A}>\gamma^{DMFT}$ for momenta on the ``Fermi surface'' (see \fref{varianceFS}),
and the spectral weight at the Fermi level is further depleted. This is the signature of the appearance of a second 
scattering mechanism ($\Im\Sigma(\vek{k}_F,\omega=0)$) in D$\Gamma$A, namely the scattering off non-local spin fluctuations.

\medskip

We further note that the large values of $Z$ at high temperatures in both DMFT and D$\Gamma$A 
are characteristic for the incoherent regime of the Hubbard model. Indeed
above a coherence temperature, $Z$ was shown to formally exceed unity in DMFT, as $\left.\partial_\omega\Re\Sigma(\vek{k},\omega)\right|_{\omega=0}$ changes to a positive sign\cite{PhysRevB.64.045103},
a very clear hallmark of non-Fermi liquid behaviour.

\section{The momentum dependence of the D$\Gamma$A self-energy: different measures}

\begin{figure*}[!th]%
\begin{tabular}{ccc}
\multicolumn{3}{l}{$\Sigma(k,\omega)=\qquad\quad\pmb{\Re\Sigma(k,\omega=0)}\qquad\qquad+\qquad\qquad\quad\left(1-1/\pmb{Z(k)}\right)\omega\qquad\qquad-\qquad\qquad\qquad\im\pmb{\Gamma(k)}\left(\omega^2+\pi^2T^2\right)\quad+\,\cdots$}\\[0.2cm]
  {\scalebox{1.}{\includegraphics[clip=true,trim=10 0 0 10,angle=-90,width=.3\textwidth]{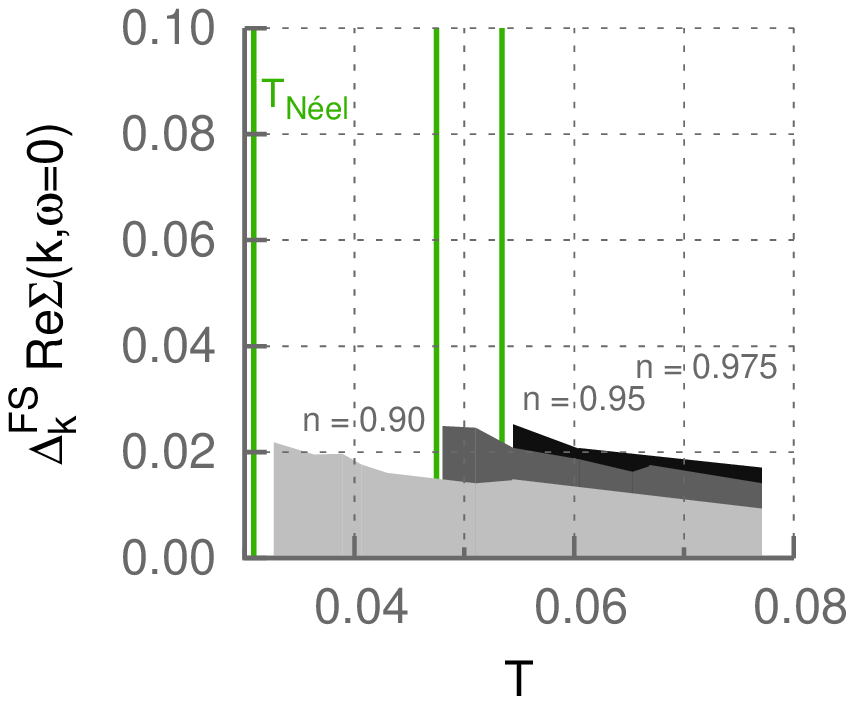}}} 
& {\scalebox{1.}{\includegraphics[clip=true,trim=10 0 0 10,angle=-90,width=.3\textwidth]{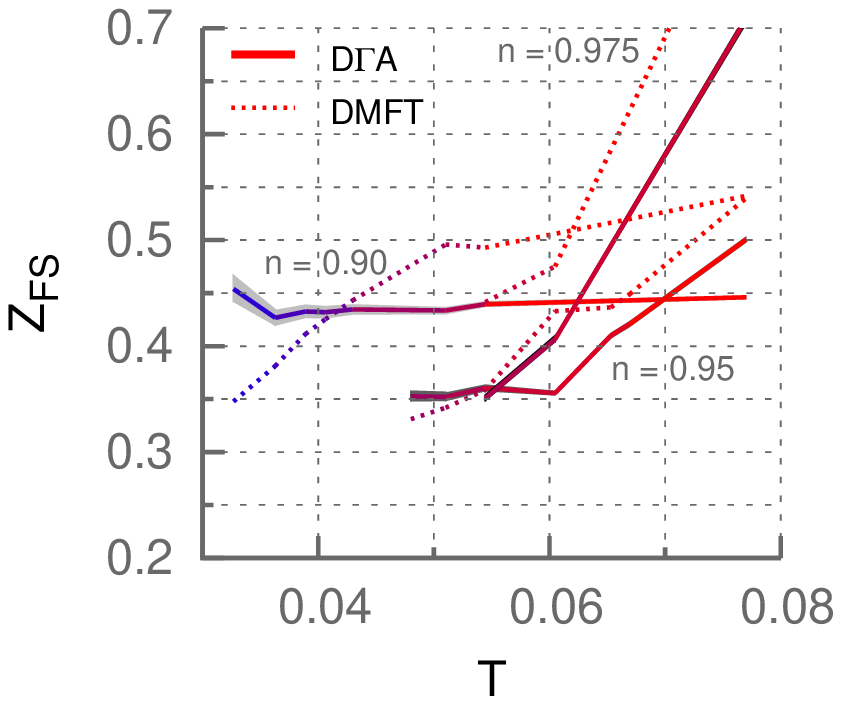}}} 
& {\scalebox{1.}{\includegraphics[clip=true,trim=10 0 0 10,angle=-90,width=.3\textwidth]{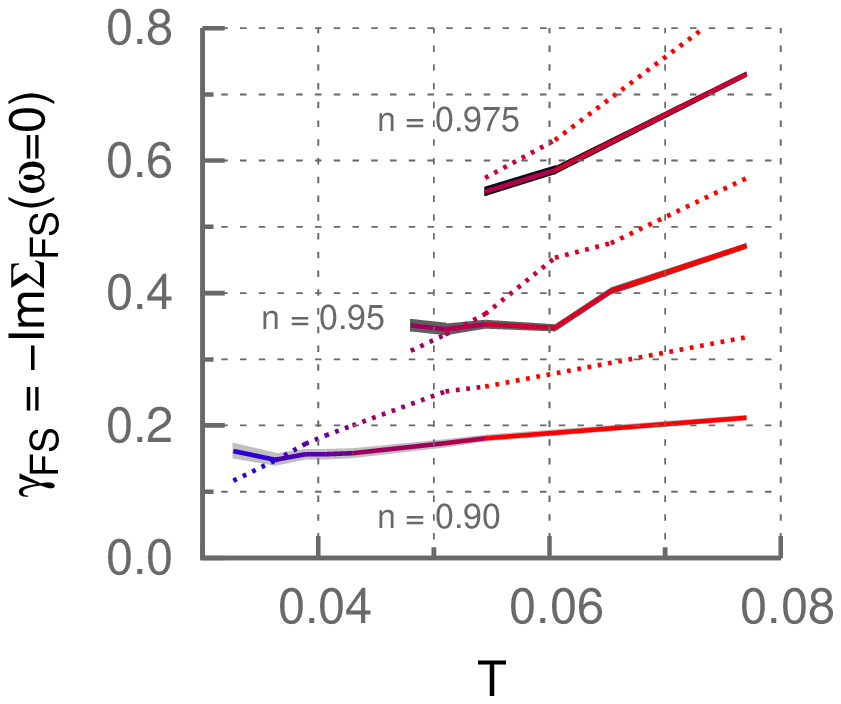}}}
\end{tabular}
\caption{
{\bf Low-energy expansion of the D$\Gamma$A self-energy and momentum dependence of the expansion coefficients {\it on the Fermi surface}.} 
This is the equivalent of Fig.\ 2 of the main manuscript, albeit with the momentum average $a_{FS}=1/N_{\svek{k}_F}\sum_{\svek{k}_F}a(\vek{k})$ and standard deviation 
restricted to the Fermi surface $\left\{\vek{k}_F\right\}$.
The shaded areas (light gray, gray, black) indicate the standard deviation, $\Delta^{FS}_{\svek{k}}a(k)=\sqrt{1/N_{\svek{k}_F}{\sum_{\svek{k}_F}\left| a(\vek{k})-a_{FS} \right|^2}}$, of the expansion coefficients $a(\vek{k})=\Re\Sigma(\vek{k},\omega$$=$$0)$, $Z(\vek{k})$, $\gamma(\vek{k})=-\Im\Sigma(\vek{k},\omega$$=$$0)=\Gamma(\vek{k})\pi^2 T^2+\mathcal{O}(T^4)$ on the Fermi surface with respect to their Fermi surface averaged values $a_{FS}$ as a function of temperature for different fillings ($n=0.9$, $n=0.95$, $n=0.975$). 
From left to right:
(i) the static real part of the self-energy at the Fermi level, $\Re\Sigma(\vek{k},\omega$$=$$0)$: Its standard deviation increases slightly on approaching the N{\'e}el transition (marked by green lines).
The variation on the Fermi surface is however found to be significantly smaller than when considering the full Brillouin zone (see Fig.\ 2 of the main manuscript).
The average value of $\Re\Sigma(\vek{k}_F,\omega$$=$$0)$, including the Hartree contribution, was absorbed into the chemical potential.
(ii) the quasi-particle weight $Z(\vek{k})$, and 
(iii) the scattering rate $\gamma(\vek{k})$. 
For (ii) and (iii) the standard deviation with respect to momentum is shown as stripes (shades of gray) around the respective Fermi surface averaged value, $Z_{FS}$ and $\gamma_{FS}$,
within D$\Gamma$A (solid lines). As comparison are shown the (by construction) local values of $Z$ and $\gamma$ within DMFT (dotted lines).
$U=1.6$ 
and the temperatures and fillings correspond to the vertical cuts shown in Fig.\ 1 in the main manuscript. 
}
\label{varianceFS}
\end{figure*}

Finally, we compare different measures for the momentum dependence of the low-energy expansion coefficients of the self-energy.
In \fref{varianceFS} we show the same data as in Fig.\ 2 of the main manuscript, albeit with the considered momenta limited to the Fermi surface:
Instead of local projections with respect to the full Brillouin zone, we sum only over momenta $\vek{k}_F$ that yield a pole in the Green's function (when neglecting life-time effects, cf.\ the comments on Luttinger's theorem above and in the main text): $a_{FS}=1/N_{\svek{k}_F}\sum_{\svek{k}_F}a(\vek{k})$.
Also the standard deviation in momentum space is limited to Fermi vectors: $\Delta^{FS}_{\svek{k}}a(k)=\sqrt{1/N_{\svek{k}_F}{\sum_{\svek{k}_F}\left| a(\vek{k})-a_{FS} \right|^2}}$.

There is a notable difference of $Z_{FS}$, $\gamma_{FS}$ to the averages over the full Brillouin zone $Z_{loc}$, $\gamma_{loc}$ (cf.\ Fig.\ 2 of the main text). On the one hand, the momentum differentiation is even more negligible when measured with $\Delta_{\svek{k}}^{FS}$ than when considering the deviation $\Delta_{\svek{k}}$ over the entire Brillouin zone.
On the other, one sees at low-$T$ a crossover to a regime with $Z_{FS}^{D\Gamma A}>Z^{DMFT}$ and $\gamma^{D\Gamma A}_{FS}>\gamma^{DMFT}$.
The behaviour in $\gamma$ is the result of the onset of non-local spin fluctuations that provide an additional scattering mechanism (in DMFT $\gamma\rightarrow 0$ for $T\rightarrow 0$%
\cite{bla}%
).
The low-$T$ rise (or at least saturation) in $Z^{D\Gamma A}_{FS}$ is in close analogy to the similar high-$T$ behaviour of $Z^{DMFT}$, which is propelled by local electron-electron scattering.

\begin{figure*}[!bh]%
\begin{tabular}{ccc}
& {\scalebox{1.}{\includegraphics[clip=true,trim=10 0 0 10,angle=-90,width=.3\textwidth]{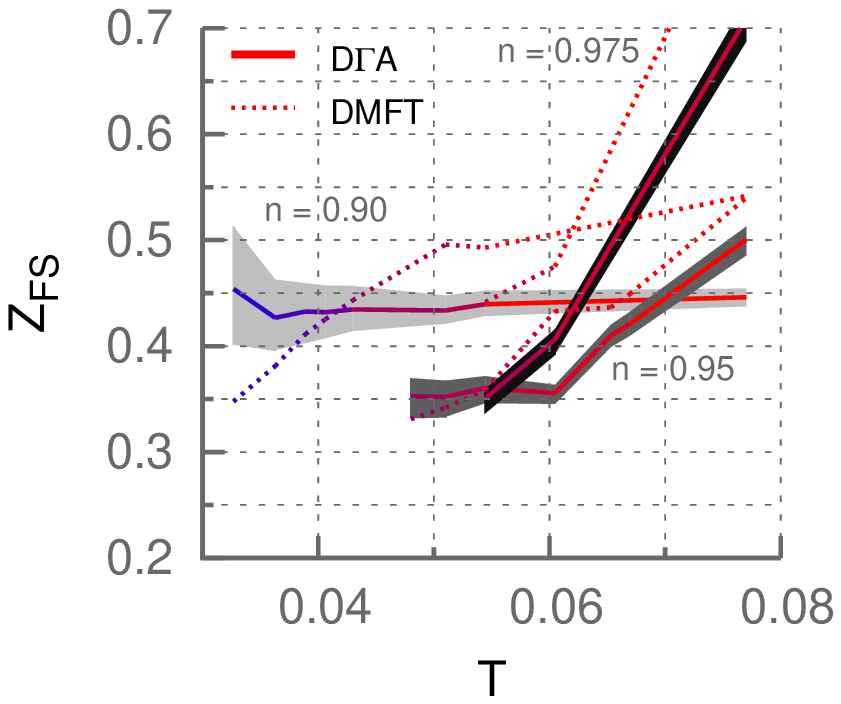}}} 
& {\scalebox{1.}{\includegraphics[clip=true,trim=10 0 0 10,angle=-90,width=.3\textwidth]{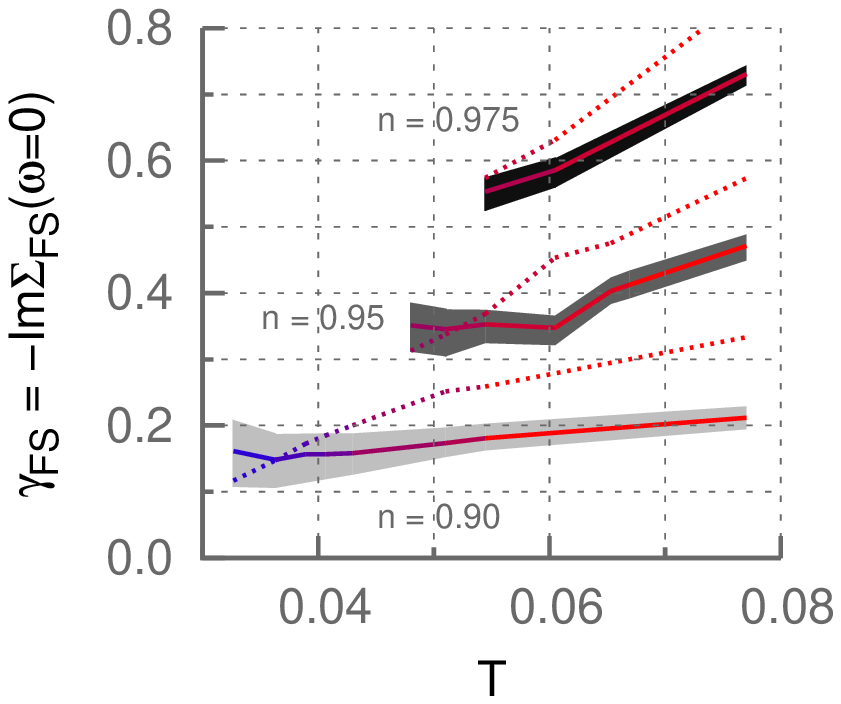}}}
\end{tabular}
\caption{
{\bf Low-energy expansion of the D$\Gamma$A self-energy and momentum dependence of the expansion coefficients {\it on the Fermi surface}.} 
This is the equivalent of \fref{varianceFS}, albeit using as measure for the momentum dependence the minimal and maximal values of $Z$ and $\gamma$ on the Fermi surface.
}
\label{varianceFS_MM}
\end{figure*}

In \fref{varianceFS_MM} we choose to represent the width of momentum dependence by the minimal and maximal values found on the Fermi surface.
Using this measure, the momentum variation is seen to increase more significantly when cooling towards the spin ordered phase.
Comparing \fref{varianceFS} and \fref{varianceFS_MM} one however realizes that the extremal $Z$'s and $\gamma$'s must be confined to only very small portions of the Fermi surface,
around $\vek{k}_2$ and $\vek{k}_3$ (see Fig.\ 3 of the main manuscript and, there, Table I).

\begin{thebibliography}{46}
\bibitem[{NoteL()}]{blaL}
\bibinfo{note}{Locality (e.g.\ in $G^{loc}$) is defined with respect to a local/Wannier basis, $\psi_{L\svek{R}}(\vek{r})$.
This is different to the COHSEX approximation\cite{hedin} 
that assumes point-locality, $\vek{r}=\vek{r}^\prime$, for correlations, and is moreover static for both exchange and correlation contributions
to the self-energy.
}
\end{thebibliography}


\end{document}